\documentclass{PoS}
\newcommand{\arxiv}[1]{arXiv:\,\href{http://arxiv.org/abs/#1}{{\tt #1}}}
\usepackage{subfigure}
\usepackage{grffile}

\title{Witten index and phase diagram of compactified $\mathcal{N}=1$ supersymmetric Yang-Mills theory on the lattice}

\ShortTitle{Witten index and phase diagram of compactified $\mathcal{N}=1$ supersymmetric Yang-Mills theory on the lattice}

\author{Georg Bergner\\
        Universität Bern, Institut für Theoretische Physik, Sidlerstr. 5, CH-3012 Bern, Switzerland\\
        E-mail: \email{bergner@itp.unibe.ch}}

\author{Pietro Giudice, Gernot M\"unster%
         \\
        Universität Münster, Institut für Theoretische Physik, Wilhelm-Klemm-Str. 9, D-48149 Münster, Germany\\
        E-mail: \email{p.giudice@uni-muenster.de}, \email{munsteg@uni-muenster.de}}
        
\author{\speaker{Stefano Piemonte}%
         \\
        Universit\"at Regensburg, Institute for Theoretical Physics, D-93040 Regensburg, Germany\\
        E-mail: \email{stefano.piemonte@ur.de}}

\abstract{Owing to confinement, the fundamental particles of $\mathcal{N}=1$ Supersymmetric Yang-Mills (SYM) theory, gluons and gluinos, appear only in colourless bound states at zero temperature. Compactifying the Euclidean time dimension with periodic boundary conditions for fermions preserves supersymmetry, and confinement is predicted to persist independently of the length of the compactified dimension. This scenario can be tested non-perturbatively with Monte-Carlo simulations on a lattice. SUSY is, however, broken on the lattice and can be recovered only in the continuum limit. The partition function of compactified $\mathcal{N}=1$ SYM theory with periodic fermion boundary conditions corresponds to the Witten index. Therefore it can be used to test whether supersymmetry is realized on the lattice. Results of our recent numerical simulations are presented, supporting the disappearance of the deconfinement transition in the supersymmetric limit and the restoration of SUSY at low energies.}

\FullConference{The 33rd International Symposium on Lattice Field Theory\\
                 14 -18 July  2015\\
                 Kobe International Conference Center, Kobe, Japan}

\begin{document}

\section{Introduction}

Confinement is the non-perturbative aspect of QCD responsible for the mass and the properties of hadrons and mesons. Perturbation theory can provide a good description of strong interactions at high energies, but a thorough analytical understanding of confinement at low energies is still missing. The 't Hooft limit \cite{Hooft1974} is a popular method to approach this problem in the context of the AdS/CFT duality. Yang-Mills theories simplify drastically if the gauge coupling $g$ is sent to zero and the number of colors $N_c$ to infinity while keeping fixed the 't Hooft coupling $\lambda = g^2 N_c$. 

For instance volume independence and correlation function factorization have been conjectured for pure Yang-Mills theories in the large $N_c$ limit. A discretized lattice model with only one site $1^d$ would be equivalent to the full theory defined in an infinite lattice \cite{Eguchi1982}. Unfortunately the hypothesis that leads to the volume reduction is not fulfilled in the large $N_c$ limit: confinement is not preserved, in the sense that the center symmetry of the Polyakov loop is spontaneously broken \cite{Bhanot1982}. A possibility to solve this issue is to couple gauge fields to quarks in the adjoint representation of the gauge group (AdjQCD), that give a stabilizing contribution to effective potential of the Polyakov loop at least at one-loop order of perturbation theory \cite{Kovtun2007,Unsal:2008ch}. If only one massless adjoint Majorana fermion is coupled to the gauge field, the model corresponds to the $\mathcal{N}=1$ Supersymmetric Yang-Mills theory and the fermionic particles, the superpartners 
of the gluons, are called gluinos.

Models with adjoint fermions are relevant not only for the full volume reduction. The \mbox{AdjQCD} theories defined in four dimensions can be compactified on a torus $\mathbb{R}^3\times S^1$ and the fermion boundary conditions in the compact dimension can be chosen to be periodic. A semiclassical expansion for small compactification radius $R$ is then expected to provide an analytical explanation for the properties of confinement also far away from the limit $N_c \rightarrow \infty$ \cite{Shifman2008,Anber2012,Anber2015a,Anber2015b}. As before, the important assumption is the absence of a deconfinement phase transition that would otherwise disconnect the small $R$ regime from the physics at zero temperature, where confinement has to be understood. Lattice Monte-Carlo simulations can provide a fully non-perturbative evidence for this scenario~\cite{Bergner2014compact}.

In this contribution we discuss our recent numerical lattice simulations for $\mathcal{N}=1$ Supersymmetric Yang-Mills (SYM) theory and we present evidence that no deconfinement phase transitions occurs in the chiral limit regardless from the length of $R$. Moreover, the partition function of compactified $\mathcal{N}=1$ SYM with periodic fermion boundary conditions corresponds to the Witten index. We show how the expected exact cancellation between bosonic and fermionic energy levels can be used as a test for the realization of supersymmetry on the lattice.

\section{The partition function and the Witten index of $\mathcal{N}=1$ SYM}

Confinement is a crucial property of the $\mathcal{N}=1$ Supersymmetric Yang-Mills (SYM) theory. At zero temperature the fundamental particles of the model, gluons and gluinos, appear only in colorless bound states. Being in the adjoint representation of the gauge group, the gluino has $N_c^2-1$ color degrees of freedom and therefore its contributions do not decouple in the large $N_c$ limit.

If one space-time dimension is compactified, antiperiodic boundary conditions applied to the gluino field $\lambda$ in the time direction bring to the trace of the Boltzmann factor $\exp(-\hat{H}R)$ in the Hamiltonian formalism:
\begin{equation}
 Z(R) = \int_{\lambda(R)=-\lambda(0)} \mathcal{D}\lambda \mathcal{D}A_\mu \exp{\left(-\int_0^{R} dt \int d^3x \mathcal{L}(\lambda,A_\mu)\right)} = \textrm{Tr}(\exp(-\hat{H}R))\,,
\end{equation}
while periodic boundary conditions provide the so-called ``supertrace'' $W(R)$:
\begin{eqnarray}
 W(R) & = & \int_{\lambda(R)=\lambda(0)} \mathcal{D}\lambda \mathcal{D}A_\mu \exp{\left(-\int_0^{R} dt \int d^3x \mathcal{L}(\lambda,A_\mu)\right)}\\ & = & \textrm{STr}(\exp(-\hat{H} R)) = \textrm{Tr}((-1)^F\exp(-\hat{H} R))\,,
\end{eqnarray}
where $F$ is the fermion number.

The Grassmanian nature of fermion fields requires antiperiodic boundary conditions in the time direction to reproduce correctly the partition function $Z(T)$ at non-zero temperature $T=1/R$ in the path-integral formalism. However $W(R)$ has an interesting cancellation between fermion and boson contributions due to the $(-1)^F$ factor, that has been useful to study many different aspects of quantum field theories. The function $W(R)$ is known for example as ``Witten index'' in SUSY and it provides information on spontaneous supersymmetry breaking \cite{Witten1982}. Furthermore, as discussed in the introduction, gluinos are expected to preserve confinement for all compactification radius $R$, if periodic boundary conditions are applied to their fields in the time direction.

We study the compactified version of $\mathcal{N}=1$ SYM on the lattice with numerical Monte-Carlo simulations for the gauge group SU(2). The integral of the continuum Lagrangian density
\begin{equation}
\mathcal{L}(\lambda, A_\mu) =- \frac{1}{4} (F_{\mu\nu}^a F_{\mu\nu}^a) + \frac{1}{2} \bar{\lambda}_a (\gamma^\mu D^{ab}_\mu + m) \lambda_b\,,
\end{equation}
can be discretized on the lattice with the tree-level Symanzik improved action and the Dirac-Wilson operator. The theory is supersymmetric in the continuum formulation if and only if the gluino mass $m$ is set to zero. Supersymmetry is however explicitly broken on the lattice and a simultaneous fine tuning of the bare gluino mass $m_B$ and gauge coupling $g$ is required to restore it in the continuum limit. The renormalized gluino mass $m_R(g,m_B)$ considered as a function of the bare parameters will be zero in general for $m_B \neq 0$. In our simulation we represent $m_R$ by the square of the adjoint pion mass in a partially quenched approach \cite{Muenster2014}.

\section{Continuity of confinement for SU(2) $\mathcal{N}=1$ SYM}

\begin{figure}
\centering
 \subfigure[Volume $8^3 \times 4$, $c_{SW}=0$]{\includegraphics[width=.47\textwidth]{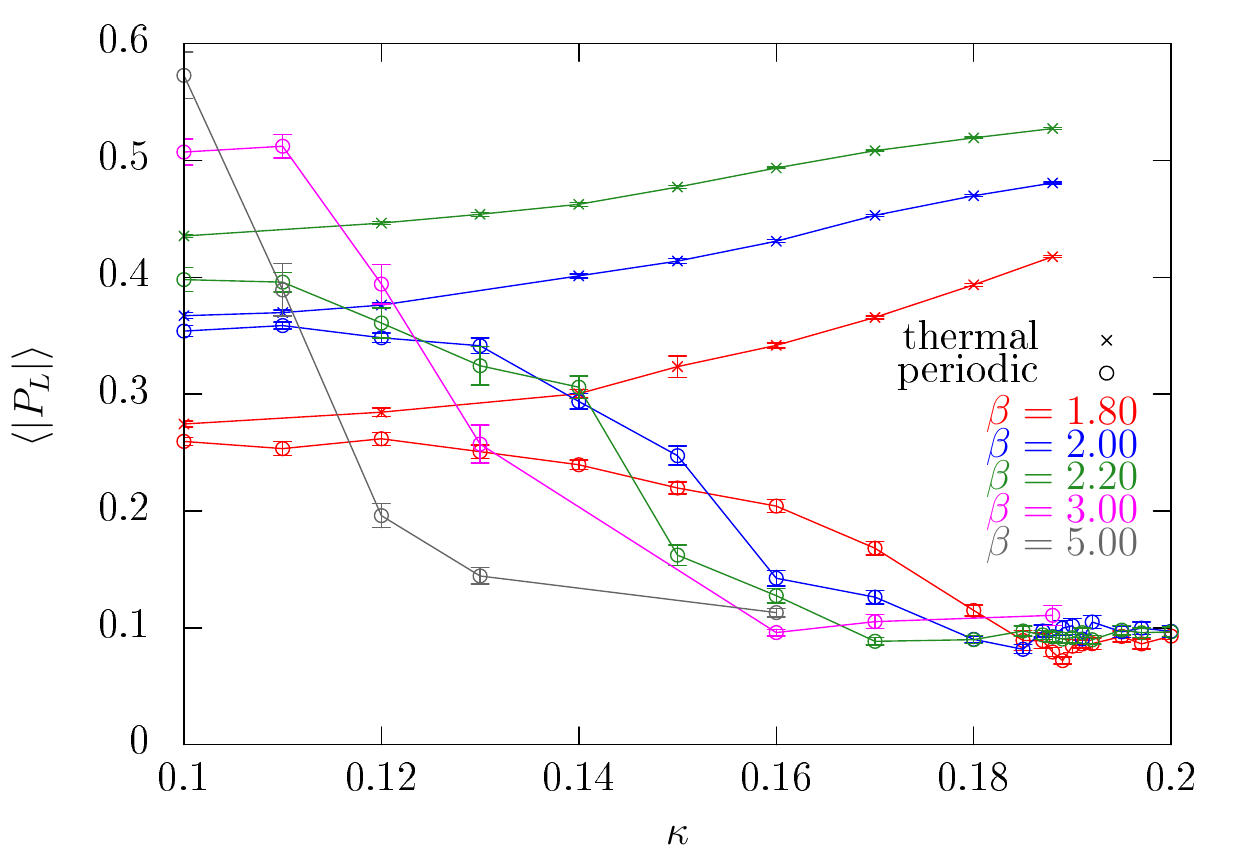}\label{plotkappascans_ns8_per_fig}}
 \subfigure[PDF for the Polyakov loop]{\includegraphics[width=.47\textwidth]{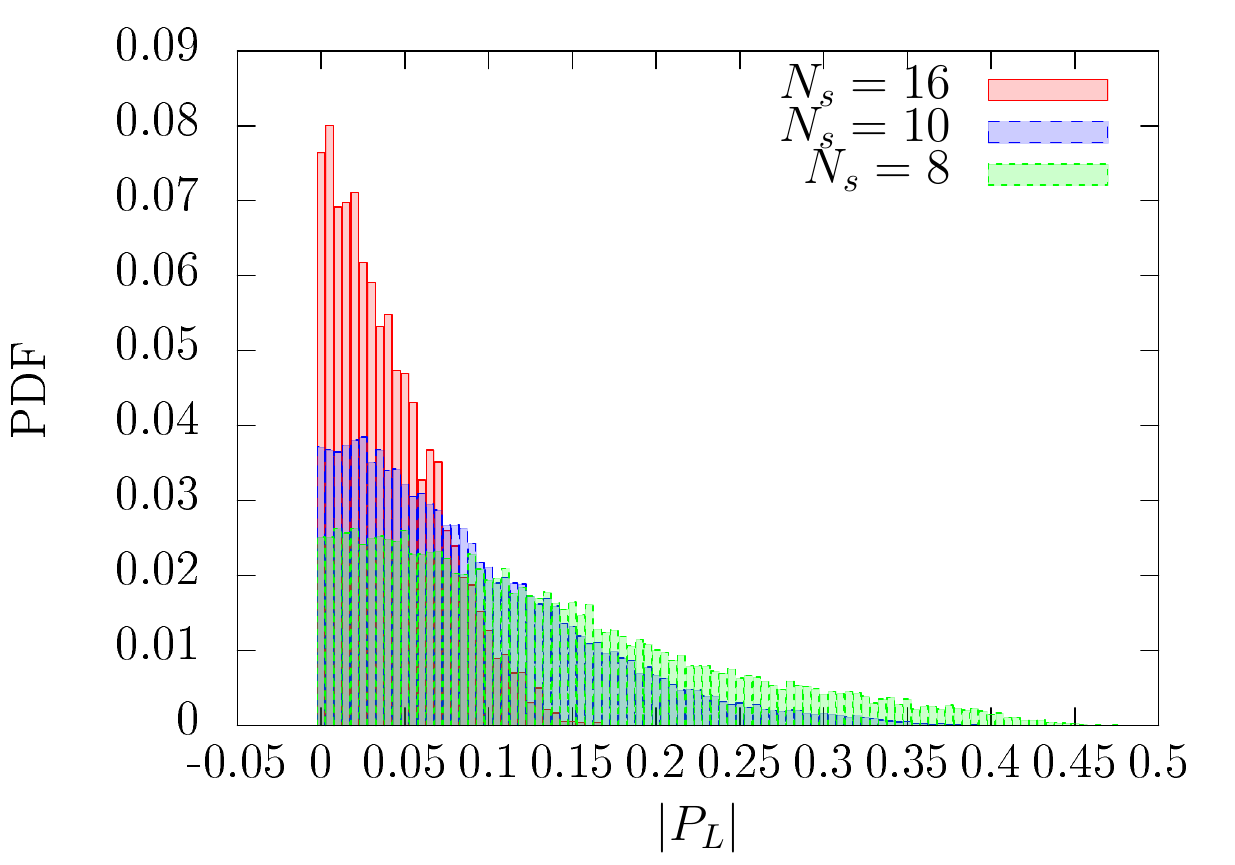}\label{finite_volume}}
 \subfigure[Observed phase diagram in the bare phase space]{\includegraphics[width=.35\textwidth]{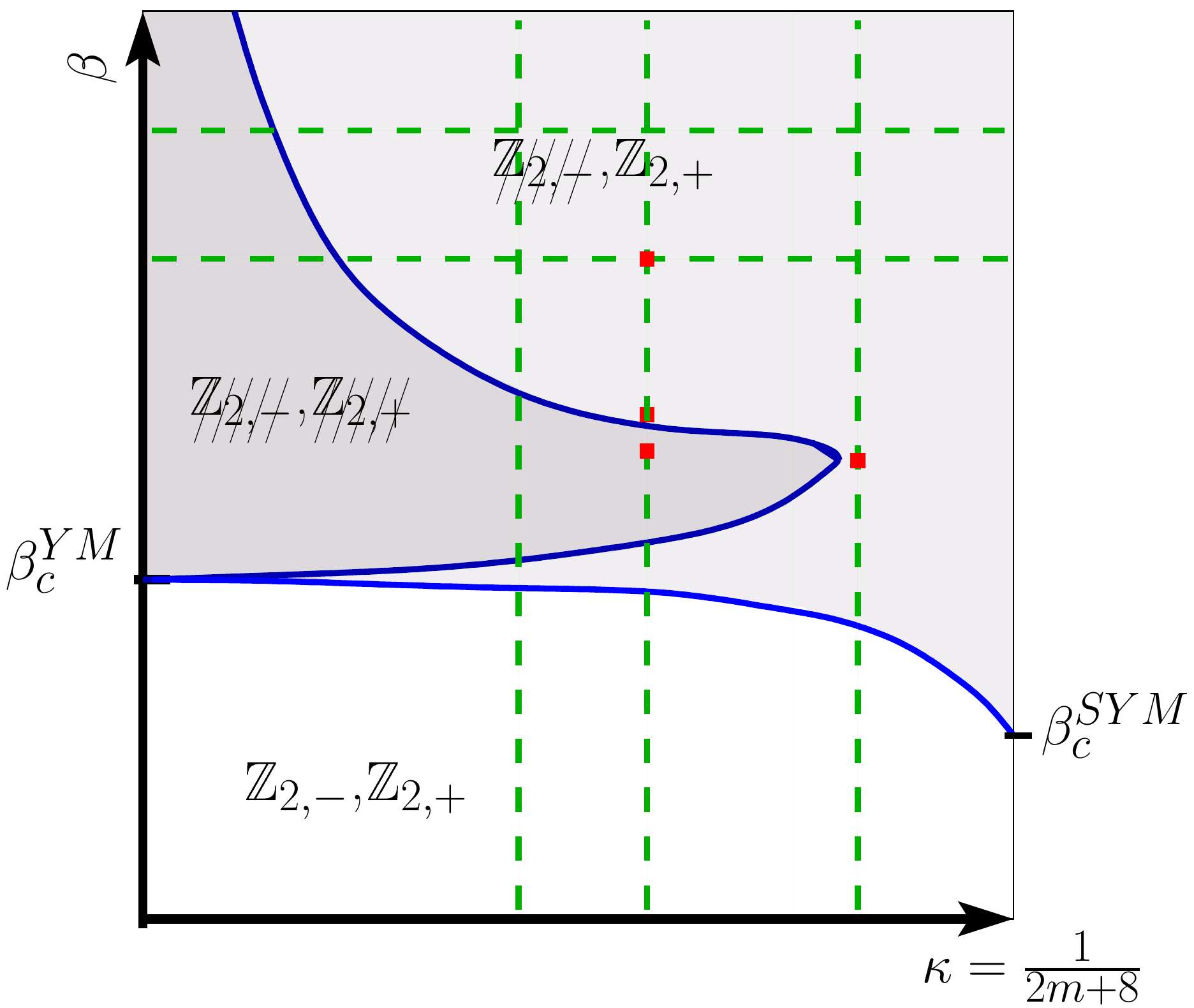}\label{diagram}}
 \subfigure[Lines of constant physics]{\includegraphics[width=.41\textwidth]{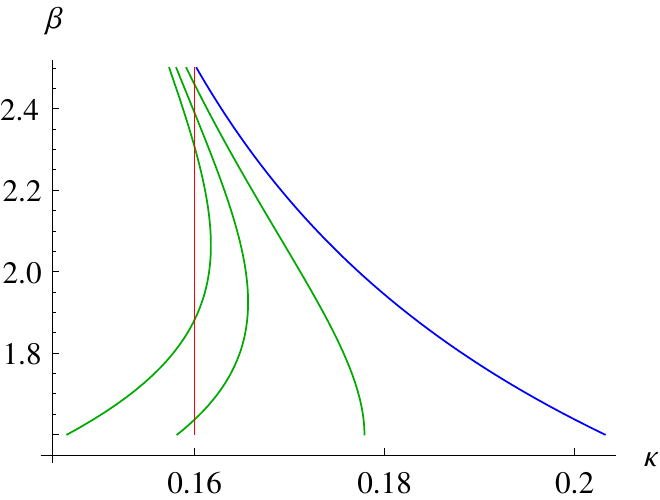}\label{lcp}}
 \caption{a) Polyakov loop expectation value on a $8^3 \times 4$ lattice as a function of $\beta$ and $\kappa$. b) Probability distribution function for the Polyakov loop as a function of the spatial volume for $\beta=2.2$ and $\kappa=0.16$. c) Phase diagram in the bare coupling space. Center symmetry is broken in the shadow region for both periodic and antiperiodic fermion boundary conditions. The lower region marked by the light blue line is instead confined for all fermion boundary conditions; in the remaining part of the coupling space confinement persists only for fermion periodic boundary conditions. d) Lines of constant physics for the unimproved Dirac-Wilson operator. The renormalized gluino mass vanishes along the blue line. Along the green lines supersymmetry is softly broken by a specific constant value of the renormalized gluino mass.}
\end{figure}
The first scan of the phase diagram has been done in the bare parameter space at fixed number of lattice points $N_t$ in the time direction. The compactification radius $R$ changes as a function of $\beta=2N_c/g^2$; in particular $R\rightarrow 0$ when $\beta\rightarrow \infty$. The results are presented in Fig.~\ref{plotkappascans_ns8_per_fig}. The expectation value of the Polyakov loop distinguishes three different regions, depending on whether center symmetry is broken for periodic or antiperiodic fermion boundary conditions, see Fig.~\ref{diagram}. Remarkably, the critical line where confinement is broken for periodic boundary conditions does not intersect the line where the gluino mass vanishes, see Fig.~\ref{lcp}, meaning that in the supersymmetric limit there are no deconfinement phase transitions, as predicted. Due to the flat behavior of the Polyakov loop effective potential for $\mathcal{N}=1$ SYM, finite volume corrections have been important for a correct identification of the phase transition point, see Fig.~\ref{finite_volume}.
\begin{figure}
\centering
\includegraphics[width=.57\textwidth]{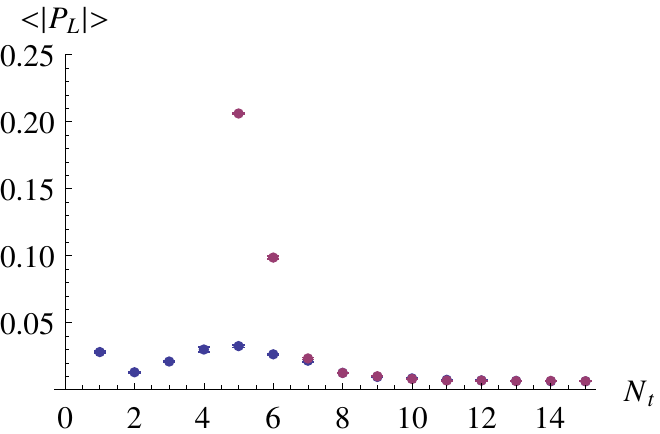}
\caption{Polyakov loop expectation value for periodic (blue) and antiperiodic (purple) boundary conditions on a $16^3\times N_t$ lattice.}\label{fixedpl}
\end{figure}
Alternatively, the couplings $\kappa$ and $\beta$ can be hold fixed and the compactification radius $R=a N_t$, where a is
the lattice spacing, can be changed by discrete steps by simply increasing and decreasing $N_t$ (\emph{fixed scale approach}). In this case the systematic uncertainties related to the determination of the line of constant physics are avoided. While the first scan has been done using unimproved Wilson fermions, this second run has been done using tree-level improved clover fermions in order to reduce the lattice artefacts \cite{Musberg2013}. We have set the volume to $16^3 \times N_t$, $\beta=1.65$, $c_{sw} = 1$ and $\kappa=0.1750$; the measured pion mass at zero temperature is $am_\pi = 0.64631(67)$. As shown in Fig.~\ref{fixedpl}, the influence of the periodic boundary conditions is drastic especially for small $R$, $\langle|P_L|\rangle$ is always bounded even for a $16^3\times1$ lattice. These results indicate further that confinement is stronger for a small than for an intermediate compactification radius.

\section{Supersymmetry and the Witten index on the lattice}

The Witten index $W(T)$ can be expressed in terms of the boson and fermion energy levels
\begin{equation}\label{wt}
W(T) = \sum_{\textrm{bosons}} \exp{\left(-\frac{E_i^B}{T}\right)} - \sum_{\textrm{fermions}} \exp{\left(-\frac{E_i^F}{T}\right)}\,,
\end{equation}
where here and in the following we set $T=1/R$ for a better comparison between the Witten index and the thermal partition function.

The Witten index is constant for a supersymmetric theory. If supersymmetry is not broken, both the number and the precise energy value of fermion and boson states will match and the above difference will be reduced to a constant equal to the number of unpaired ground states. It is for example well known that $W(T)$ is equal to $N_c$ for the $\mathcal{N}=1$ SYM theory \cite{Witten1982}. Supersymmetry is however explicitly broken on the lattice: the continuous translational symmetry is reduced to the discrete subgroup of translations proportional to multiples of the lattice spacing $a$. In addition, for $\mathcal{N}=1$ SYM theory, numerical simulations are impossible at exactly vanishing gluino mass. 

A computation of $W(T)$ or equivalently of the difference of all bosonic and fermionic energy levels would be a precise measure of the supersymmetry breaking on the lattice. This approach is however rather difficult to implement since it would imply the knowledge of the value of the path integral with its absolute normalization factor. However supersymmetry is explicitly broken on the lattice, and therefore $W(T)$ is not anymore a constant independent from $T$ since $E_i^B \neq E_i^F$. If we derive the expression \ref{wt} above
\begin{equation}
E_G(T) = \frac{\partial W(T)}{\partial (1/T)} = \sum_{\textrm{fermions}} E_i^F \exp{\left(-\frac{E_i^F}{T}\right)} - \sum_{\textrm{bosons}} E_i^B \exp{\left(-\frac{E_i^B}{T}\right)}
\end{equation}
and subtract a zero temperature value, $E_G(T) - E_G(0)$, we get a \emph{graded energy density} that can be easily computed on the lattice. For unbroken supersymmetry this quantity should be zero.  Any deviation of $E_G(T)$ from zero would then be a signal of explicit supersymmetry breaking. The graded energy density $E_G(T)$ becomes sensible to the mismatch of larger and larger fermion and boson energy states at $T$ increases, i.e. when the compactification radius $R$ decreases.

\begin{figure}
\centering
\includegraphics[width=.55\textwidth]{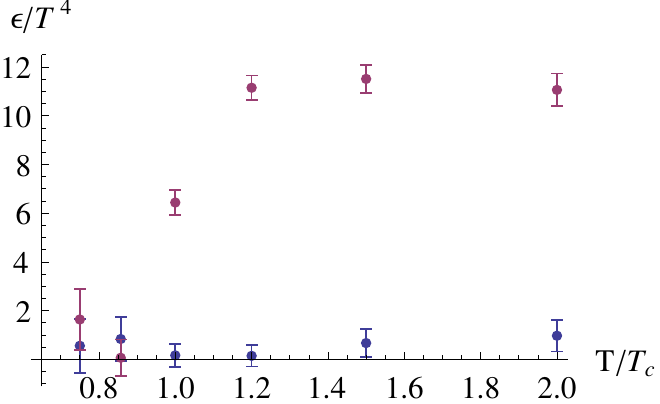}
\caption{Energy density $\epsilon=E_G(T)/V$ for periodic (blue) and antiperiodic (purple) gluino boundary conditions.}\label{ppthermo}
\end{figure}
To compute $E_G(T)$ on the lattice, we simulate the theory for different $N_t$ with periodic fermion boundary conditions, then we proceed to the calculation of the energy density as in standard thermodynamics, see for example Ref.~\cite{Ejiri2012}. The final result is shown in Fig.~\ref{ppthermo}. The energy density $\epsilon$ for antiperiodic boundary conditions shows the expected thermal behavior, $\epsilon$ is small before that the deconfinement phase transition occurs and it grows as $T^4$ at high temperature as for a free gas of gluons and gluinos. The graded energy density is instead compatible with zero even for compactification radius $R$ twice smaller the critical $R_c=1/T_c$ where the deconfinement phase transition occurs with fermion antiperiodic boundary conditions. appears only for a lattice with two sites in the temporal direction. In that sense
the critical compactification length $L = 3a$ is the minimal distance where supersymmetry is approximatively and dynamically restored in the infrared regime. Strong lattice artefacts appear at distances smaller than or equal to two lattice spacings.

It is finally interesting to observe that $E_G(T)$ exhibits a flat behavior for rather large values of the adjoint pion mass, $am_\pi \simeq 0.6$. This fact is in agreement with the previous investigations of the bound spectrum done by the DESY-M\"unster collaboration, where a degeneracy between the gluino-glue and its boson counterpart has been observed to persist even for large value of the adjoint pion mass, while a large lattice spacing leads to strong mass splitting \cite{Bergner:2012rv,Bergner:2013nwa}.

\section{Conclusions}

We have presented the first evidence for the persistence of confinement in $\mathcal{N}=1$ supersymmetric Yang-Mills theory for all compactification radius $R$ if periodic boundary conditions are applied to the gluino field in all directions. In the near future we plan new simulations for different number of colors and for different numbers of Majorana fermions $N_f > 1$, to extend the results of Ref.~\cite{Cossu2009}.

\section{Acknowledgements}

We thank I. Montvay for helpful instructions and comments. The authors gratefully acknowledge the computing time granted by the Leibniz-Rechenzentrum (LRZ) in M\"unchen provided on the supercomputer SuperMUC. Further computing time has been provided by the cluster PALMA of the University of M\"unster.

\end{document}